\documentclass{article}
\usepackage[dvips]{epsfig}

\usepackage{color}
\usepackage{amsmath}
\usepackage{amssymb}

\title{Introduction of longitudinal and transverse Lagrangian velocity increments in homogeneous and isotropic turbulence}

\author{Emmanuel L\'ev\^eque$^{(1)}$  Aurore Naso$^{(1)}$ \\              
	(1) Laboratoire de M\'ecanique des Fluides et d'Acoustique, \\ \'Ecole Centrale de Lyon \& CNRS, \\
	Universit\'e de Lyon, F-69134 \'Ecully cedex, France \\
}

\begin{document}
	\maketitle
	
\abstract{
Based on geometric considerations, longitudinal ($\delta u_\parallel^{(L)}(\tau)$) and transverse ($\delta u_\perp^{(L)}(\tau)$) Lagrangian velocity increments are introduced as components along, and perpendicular to, the displacement of fluid particles during a time scale $\tau$. It is argued that these two increments probe preferentially the stretching and spinning of material fluid elements, respectively. This property is confirmed (in the limit of vanishing $\tau$) by examining the variances of these increments conditioned on the local topology of the flow. 
These longitudinal and transverse Lagrangian increments are found to share some qualitative features with their Eulerian counterparts. 
In particular, direct numerical simulations at $R_\lambda$ up to $300$ show that the distributions of $\delta u_\parallel^{(L)}(\tau)$ are negatively skewed at all $\tau$, which is a signature of time irreversibility in the Lagrangian framework. 
Transverse increments are found more intermittent than longitudinal increments, as quantified by the comparison of their respective flatnesses and scaling laws.  
Although different in nature, standard Cartesian Lagrangian increments (projected on fixed axis) exhibit scaling properties that are very close to transverse Lagrangian increments. 
}

\section{Introduction}

Following Kolmogorov's seminal work in 1941, fluid turbulence has been extensively studied in the Eulerian framework with a focus on spatial velocity increments $\delta {\bf u}^{(E)}({\bf x}|{\bf r},t)={\bf u}({\bf x}+{\bf r},t)-{\bf u}({\bf x},t)$ \cite{mcComb}. 
In (stationary) homogeneous and isotropic (HI) turbulence, 
the statistics of  $\delta {\bf u}^{(E)}({\bf x}|{\bf r},t)$ depends only on the separation scale $r=\|{\bf r}\|$ and  
$\delta {\bf u}^{(E)}({\bf x}|{\bf r},t)$ can be profitably projected onto preferential directions along and perpendicular to 
$\hat{\bf r}={\bf r}/r$, thus defining the (scalar) longitudinal and transverse 
increments:
\begin{eqnarray}
 \delta u_\parallel^{(E)}(\bf {x}|{\bf r},t)&=&\delta {\bf u}^{(E)}({\bf x}|{\bf r},t)\cdot\hat{\bf r} 
 \label{eq:dvElong} \\
 \delta u_\perp^{(E)}({\bf x}|{\bf r},t)&=&\|P ({\bf \hat r})\cdot \delta {\bf u}^{(E)}({\bf x}|{\bf r},t)\|\times\cos\theta 
\end{eqnarray} 
 where $P_{ij}({\bf \hat r}) = \delta_{ij}-\hat{r}_i\hat{r}_j$ is a projection in the plane perpendicular to $\hat{\bf r}$ and 
 $\theta$ is a random angle in $[0,2\pi[$. 
%
%
The interest in the longitudinal increment is reinforced by the Kolmogorov's 4/5 law  
\begin{equation} \left < {\delta u_\parallel^{(E)}}^3 ({\bf x}|{\bf r},t) \right > =-\frac 4 5\varepsilon~ r,
\label{eq:4/5law}
\end{equation}
which establishes an exact equation (at inertial scales $r$) for the third-order moment, where $\varepsilon>0$ is the mean rate of  energy dissipation. The resulting negative skewness of the probability distribution function (PDF) of $\delta u_\parallel^{(E)}(r)$ signifies the time irreversibility of turbulence dynamics (at scale $r$) in the Eulerian framework \cite{frisch}.

Alternatively, there has been a growing interest in the last decades in examining turbulence from a Lagrangian viewpoint, \textit{i.e} by tracing statistical correlations along trajectories of fluid particles (see \cite{review-toschi} for a review).  
In that case, velocity increments are readily transposed as  $\delta \mathbf{u}^{(L)}(\mathbf{x},t|s) = \mathbf{u}(\mathbf{x},t|s) - \mathbf{u}(\mathbf{x},t|t)$ where $\mathbf{u}(\mathbf{x},t|s)$ denotes the velocity at time $s$ of a fluid particle that passed through the position $\mathbf{x}$ at time $t$ \cite{kraichnan}. 
Existing studies have mainly focused on the statistical description of ``Cartesian'' increments defined as the projection of $\delta \mathbf{u}^{(L)}(\mathbf{x},t|s)$ on a (fixed) Cartesian coordinate frame, \textit{i.e.} $\delta u_\mathrm{x,y,z}^{(L)}(\mathbf{x},t|s) = \delta \mathbf{u}^{(L)}(\mathbf{x},t|s)\cdot \mathbf{e_{\mathrm{x,y,z}}}$. Under the assumption of isotropy, the $x$, $y$ and $z$-increments are statistically equivalent. 
Some interest has been taken  in particular to establish a {formal} link between Eulerian and Lagrangian increments via the multifractal formalism  \cite{schmitt,Benzi-lagrangian,chevillard-CR} or transition probabilities \cite{Kamps09,Homann09}.
 In the following, a novel decomposition of  $\delta \mathbf{u}^{(L)}(\mathbf{x},t|s)$ in terms of longitudinal and transverse  increments is introduced as a natural extension of their Eulerian counterparts.

\section{New Lagrangian velocity increments}

\begin{figure}[t]
	\scalebox{0.8}{\input{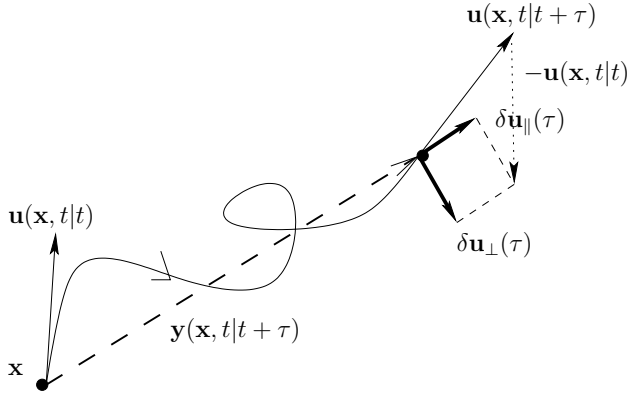}}
	
	\caption{Sketch of the longitudinal and transverse Lagrangian velocity increments, along and perpendicular to, the direction given by the displacement vector.}
	\label{fig:schema}
\end{figure}

{The ability to carry and distort material fluid elements into intricate geometries is a striking feature of turbulence. There is a consensus that statistics should connect explicitly to these peculiar geometric properties. In this respect, significant  insights have been obtained  from simplified (mathematically tractable) models of the Navier-Stokes equations, through which major trends can be related (within a Lagrangian approach) to the self-stretching and rotation of the velocity-gradient tensor, or from multi-particle velocity differences, in a local coordinate frame \cite{pumir,meneveau,chevillard,PN10}. Our work is  in line with these works with an interest in Lagrangian correlations along single-particle trajectory.}

The acceleration of a material point (or particle) is usually decomposed into a tangential and a normal component. The tangential acceleration quantifies the variation of the magnitude of the velocity (and therefore relates to the variation of kinetic energy of the particle) whereas the normal acceleration is sensitive to the curvature of the trajectory. In our context, it is natural to seek for a similar decomposition for the Lagrangian velocity increment $\delta \textbf{u}^{(L)}(\mathbf{x},t|s)$. Accordingly, it is proposed to split $\delta \textbf{u}^{(L)}(\mathbf{x},t|s)$ into a longitudinal and a transverse component, along and perpendicular to the direction indicated by the overall displacement $\mathbf{y}(\mathbf{x},t|s)=\int_t^s \mathbf{u}(\mathbf{x},t|s^\prime) ds^\prime$ (see Fig. \ref{fig:schema}). This splitting somewhat generalizes the decomposition of the (instantaneous) acceleration to the coarse-grained   dynamics at time scale $\tau$. The (scalar) longitudinal increment can  be directly identified as
\begin{equation}
\delta u_\parallel^{(L)}(\mathbf{x},t|s) = \delta{\bf u}^{(L)}(\mathbf{x},t|s) \cdot \hat{\bf y}(\mathbf{x},t|s), \label{eq:lagrangian_increment_longitudinal}
\end{equation}
with $\hat{\mathbf{y}} =\mathbf{y}/\|\mathbf{y}\|$. This definition is formally equivalent to the Eulerian longitudinal increment Eq.(\ref{eq:dvElong}) except that the separation scale is now given by the displacement of the fluid particle during the time interval $\tau=s-t$. Similarly, the (scalar) transverse increment writes
\begin{equation}
{\delta u_\perp^{(L)}}(\mathbf{x},t|s) =  \| P(\hat{\bf y})\cdot {\delta {\bf u}}^{(L)}(\mathbf{x},t|s)\| \times \cos \theta 
\label{eq:lagrangian_increment_transverse}
\end{equation}
where $\theta$ is an independent random angle uniformly distributed in $[0,2\pi[$ (in isotropic turbulence). In brief, these increments may be viewed as an extension of Eulerian increments, obtained by replacing the fixed separation $\bf r$ by the moving displacement ${\bf y}({\bf x},t|s)$ in their definition. 
To be physically relevant, the displacement should be taken in a coordinate frame attached to the (uniform) mean flow (in HI turbulence).  
If turbulence is stationary, the statistics of $\delta u_\parallel^{(L)}(\mathbf{x},t|s)$ and $\delta u_\perp^{(L)}(\mathbf{x},t|s)$ only depend on the time scale $\tau$.

\begin{figure}[t!]
  \centering
	  \includegraphics[clip,height=6.3cm,keepaspectratio]{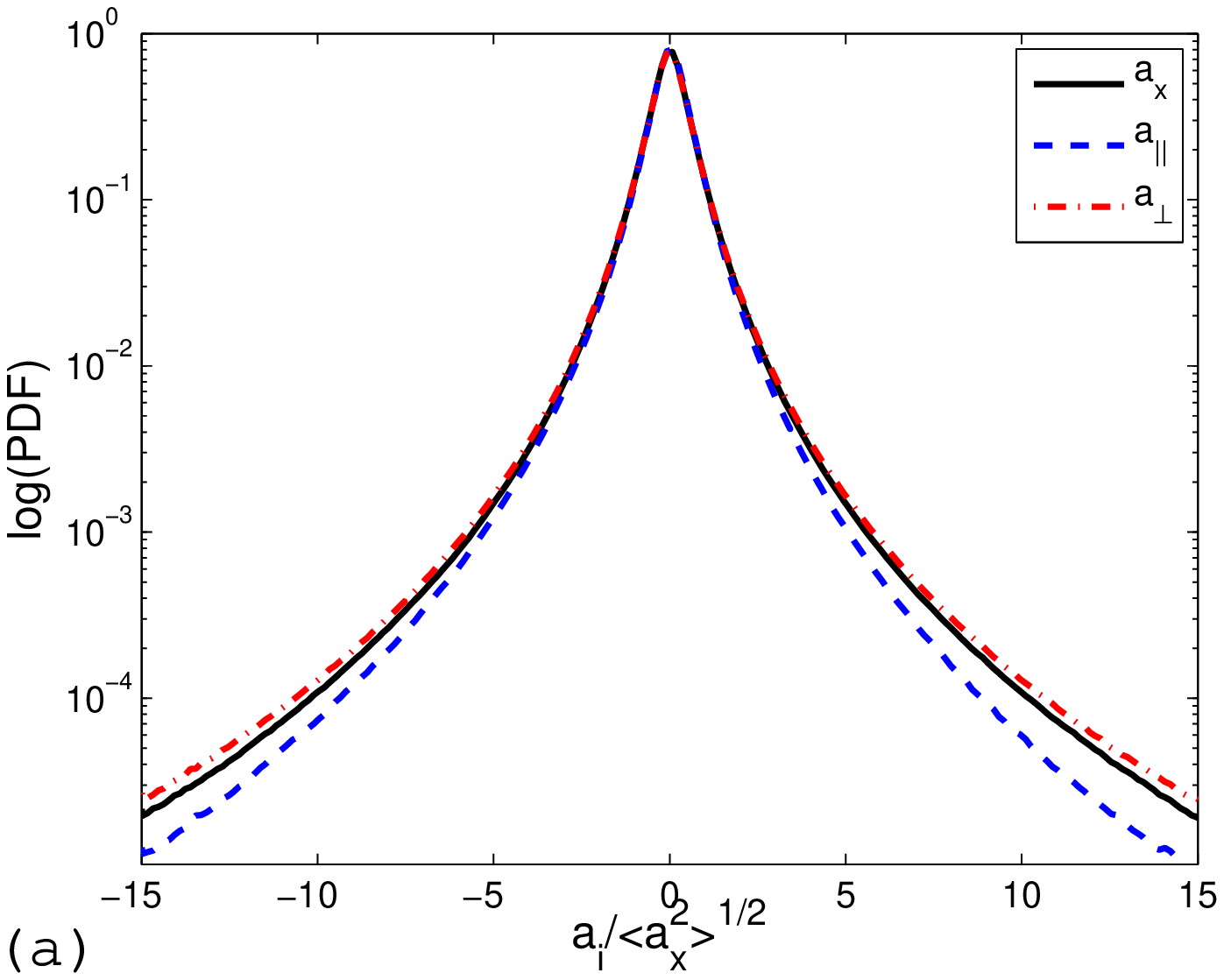}
	  \includegraphics[clip,height=6.3cm,keepaspectratio]{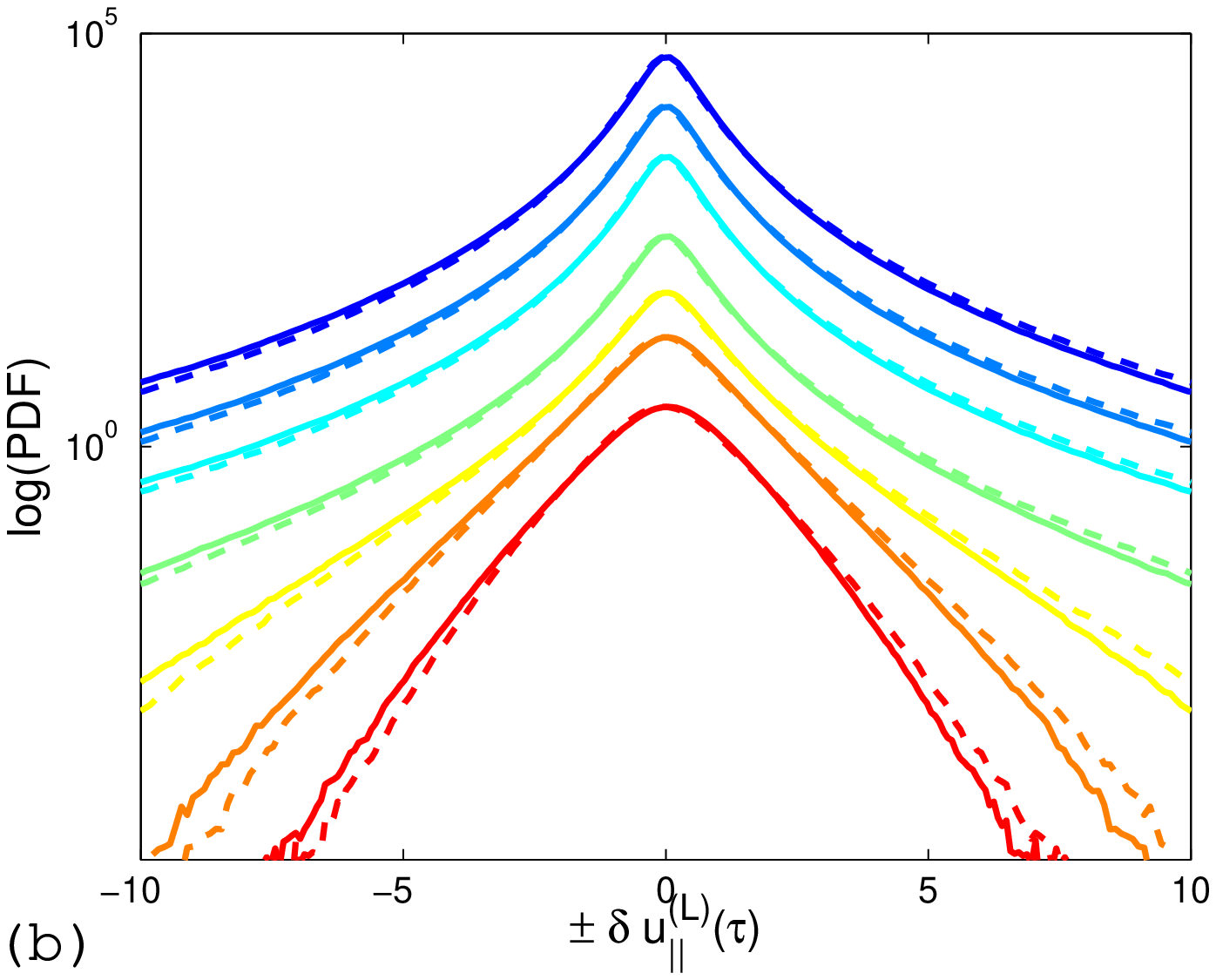}
\caption{ (a) PDFs of components of the fluid acceleration normalized by the rms value of the Cartesian acceleration $\langle a_{x}^2\rangle^{1/2}$. (b) PDFs of $\pm \delta u_\parallel^{(L)}(\tau)$ for $\tau/\tau_\eta=2.6.10^{-2},0.11,0.42, 3.4, 6.8, 14, 27$ from top to bottom. Solid lines are for $P(\delta u^{(L)}_\parallel)$ and dashed lines for $P(-\delta u^{(L)}_\parallel)$.  The curves have been vertically shifted for clarity. The turbulent Reynolds number $R_\lambda=280$.}
	\label{fig:PDF}
\end{figure}

\section{Numerical computations}

The statistics of Lagrangian velocity  increments has been investigated by pseudo-spectral (de-aliased) direct numerical simulation  (DNS) of the Navier-Stokes equations in a cubic box of size $2\pi$ with periodic boundary conditions in all directions and grid resolutions $N^3 =256^3$, $512^3$ and $1024^3$, corresponding to Reynolds numbers $R_\lambda=130$, $180$ and $280$ respectively. Time marching is operated by a second-order Adams-Bashforth scheme. An external force acts on low-wavenumber modes (at $k<2.5$)  to ensure a constant injection rate of energy, $\epsilon$,  and reach stationary HI turbulence \cite{Lamorgese}. In each simulation,  $\epsilon = 10^{-3}\mathrm{m^2.s^{-3}}$ and the viscosity $\nu$ has been adjusted so that the Kolmogorov's scale $\eta=(\nu^3/\varepsilon)^{1/4}$ remains comparable to the grid resolution $\Delta x = 2\pi/N$: $\Delta x/\eta \approx 1.5$ in agreement with standard requirements for DNS and  particle tracking \cite{calzavarini}. Fluid-particle trajectories have been integrated by using a second-order Runge Kutta scheme and a Verlet velocity algorithm. The velocity of particles has been estimated by resorting to tricubic interpolation. Statistics relies on the tracking of $48^3$, $64^3$ and $100^3$ particles respectively (uniformly distributed at initial time) during about $10$  eddy-turnover times, therefore ensuring a (checked) satisfactory statistical convergence. 
{The Kolmogorov's time scale $\tau_\eta=(\nu/\varepsilon)^{1/2}$ is used as the reference Lagrangian time scale for each simulation. The Lagrangian integral time scale, $T_L$, verifies $T_L/\tau_\eta \approx 13$, $18$ and $29$ at $R_\lambda=130$, $180$ and $280$ respectively; these values are marked by arrows on the figures.}

The PDFs of the Cartesian, longitudinal and transverse (scalar) Lagrangian increments at $R_\lambda=280$ are compared in Fig. \ref{fig:PDF}(a) as $\tau\rightarrow 0$; a limit in which velocity increments reduce to the components of the acceleration (ignoring the multiplicative factor $\tau$):  $\delta u^{(L)}_i(\mathbf{x},t|t+\tau) \approx a_i(\mathbf{x},t|t) \cdot \tau$
where $i$ denotes either $x$, $\parallel$ or $\perp$. 
All increments exhibit usual trends of acceleration PDFs with very large tails  \cite{review-toschi}.
Nevertheless, one can point out 
 that (i) the variances verify $\langle a_\parallel^2 \rangle < \langle a_{x}^2 \rangle \lesssim \langle a_\perp^2 \rangle$ with the PDFs of $a_x$ and $a_\perp$ being very close to each other, and that (ii) the PDF of $a_\parallel$ is negatively skewed. 
The ordering of the variances remains valid at all time scales (not shown):
\begin{equation}
\left < {\delta u^{(L)}_\parallel(\tau)}^2 \right > < \left < {\delta u^{(L)}_{x}(\tau)}^2 \right > \lesssim \left < {\delta u^{(L)}_\perp(\tau)}^2 \right >,  
~~\forall \tau>0.
\label{ordre_2}
\end{equation}
Interestingly, $\langle \delta u^{(L)}_\parallel(\tau)^2 \rangle < \langle \delta u^{(L)}_\perp(\tau)^2 \rangle$ is similar to the ordering verified by the longitudinal and transverse Eulerian increments. 
However, these latter quantities behave differently since $\langle \delta u_\parallel^{(E)}(r)^2 \rangle / \langle \delta u_\perp^{(E)}(r)^2 \rangle  \rightarrow 1$ as $r\rightarrow 0$, which is not satisfied by the Lagrangian increments as $\tau \to 0$.  
Indeed, the fluctuations of the normal component of the acceleration are  more intense than the longitudinal ones.
From the definitions and by assuming isotropy, $3 \langle \delta u^{(L)}_{x}(\tau)^2 \rangle = \langle \delta u^{(L)}_\parallel(\tau)^2 \rangle +2 \langle \delta u^{(L)}_\perp(\tau)^2 \rangle$. This equation enforces that $\langle \delta u^{(L)}_{x}(\tau)^2 \rangle$ must be bounded by the two other variances, which is indeed satisfied.

\begin{figure}[t!]
	\centering
	\includegraphics[clip,height=6.3cm,keepaspectratio]{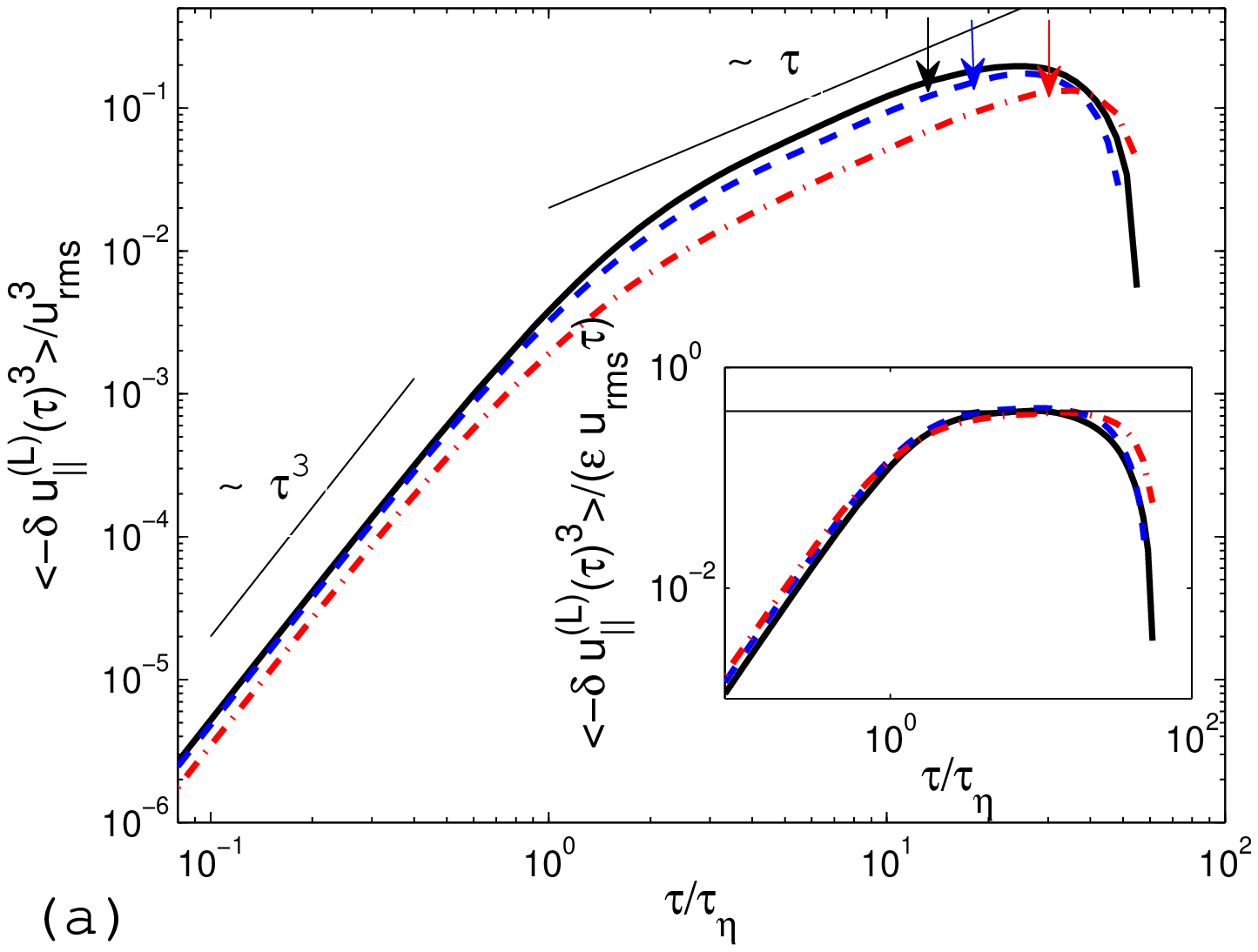}
	\includegraphics[clip,height=6.3cm,keepaspectratio]{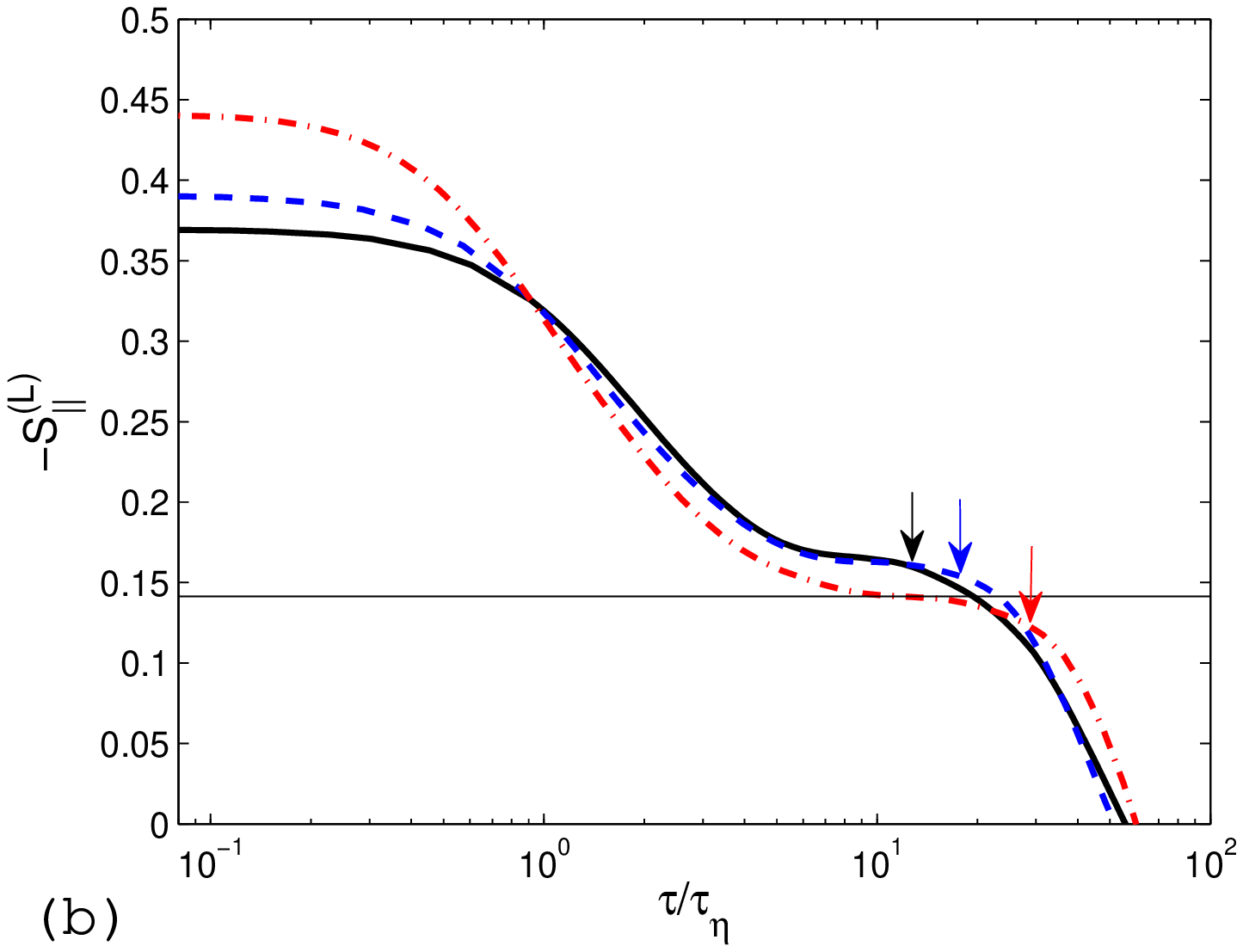}
	\caption{ Dependence on $\tau$ of (a) the third-order moment of the longitudinal Lagrangian increment -- Inset: compensated by $\varepsilon u_\mathrm{rms} \tau$ with $u_\mathrm{rms} = \langle u_x^2  \rangle^{1/2}$; (b) the skewness coefficient $S^{(L)}_\parallel(\tau)$. The horizontal lines indicate $2/5$ in the inset of (a) and $1/5\sqrt{2}\approx 0.14$ in (b). Solid line (black): $R_\lambda=130$; dashed line (blue): $R_\lambda=180$; dash-dotted line (red): $R_\lambda=280$. Arrows mark the Lagrangian integral scales $T_L$.}
	\label{fig:3}
\end{figure}

The negative skewness of the longitudinal acceleration pertains at all time scales $\tau$ for the longitudinal Lagrangian increment, as seen in Fig. \ref{fig:PDF}(b). This remarkable feature is confirmed by the computation of the skewness coefficient $S_\parallel^{(L)}(\tau)$ as a function of $\tau$ (see Fig.\ref{fig:3}):
\begin{equation}
S^{(L)}_\parallel(\tau) \equiv \frac{\left < \delta u^{(L)}_\parallel (\tau)^3 \right >}{\left < \delta u^{(L)}_\parallel (\tau)^2 \right >^{3/2}} < 0, \quad \forall \tau>0.
\label{eq:negative-skewness}
\end{equation}
This negative skewness signifies the time irreversibility of the turbulence dynamics, which is one of the most striking features of turbulence 
\cite{kraichnan-statistical-mechanics}.
In the Eulerian framework, time irreversibility is also revealed by the negative skewness of longitudinal increments (as reminded in the introduction) which points out another similarity between Eulerian and Lagrangian increments. However, there is here no exact derivation of an equation similar to Eq. (\ref{eq:4/5law}), the main difficulty arising from the pressure-gradient term, which cannot be eliminated easily in a Lagrangian coordinate system \cite{kraichnan}.

In the Lagrangian framework, time irreversibility has been first evidenced by considering at least two distinct Lagrangian particles \cite{FGV,PSC}. Indeed, any correlation function invariant under Galilean transformation and involving only one Lagrangian particle is necessarily invariant under the time reversal $\tau\rightarrow -\tau$ in a statistically homogeneous and stationary flow and, therefore, cannot discriminate time irreversibility \cite{falkovich12}.
For instance, irreversibility can not be captured by the Cartesian increment $\delta u^{(L)}_{x}(\tau)$, which is Galilean invariant \cite{mordant,Xu06}. It can not be captured by the transverse increment either, since its statistics is invariant under the change $\tau\rightarrow -\tau$ according to Eq. (\ref{eq:lagrangian_increment_transverse}).

From a kinematic viewpoint, $a_\parallel$ represents the rate of change of the amplitude of the velocity: $a_\parallel = {\mathrm{d}|{\bf u}|}/{\mathrm{d}t}$. Therefore, the negative skewness of $a_\parallel$ indicates that a fluid particle undergoes, in average, stronger deceleration than (positive) acceleration; a property that has been very recently highlighted as ``flight-crash events" in turbulence \cite{Xu14}. This feature remains valid for coarse-grained Lagrangian dynamics, if one assumes that $\delta \mathbf{u}^{(L)}(\tau)/\tau$ represents a coarse-grained acceleration at scale $\tau$.  
One should also mention that time irreversibility has  been identified by considering the skewness of the PDF of the power received by a fluid particle along its trajectory, $p={\bf a}\cdot{\bf u}$, or by considering the Lagrangian increments of the kinetic energy  \cite{mordant_th,Xu14}.
Among these multiple signatures of time irreversibility,
 our proposal has the merit to connect it to the classical phenomenology of turbulence, essentially based on the consideration of velocity increments.

A more quantitative analysis of longitudinal and transverse Lagrangian increments can be achieved by investigating the dependence in $\tau$ of their (first) statistical moments. The third-order moment of the longitudinal increment is shown in Fig. \ref{fig:3}(a) for different $R_\lambda$. 
In the dissipative range ($\tau <\tau_\eta$): $\langle \delta u^{(L)}_\parallel (\tau)^3 \rangle \sim \tau^3$ as expected. 
{At larger $\tau$, $\langle \delta u^{(L)}_\parallel (\tau)^3 \rangle$ displays a 
scale dependence that is \textit{close} to the power law $\tau$. A quantitative estimation of the scaling exponent at the inflexion point of the local slope would rather yield 
$ \tau^{1.09 \pm 0.06}$ at $R_\lambda=280$. 
In the inset of  Fig. \ref{fig:3}(a), $\langle \delta u^{(L)}_\parallel (\tau)^3 \rangle / \varepsilon u_\mathrm{rms} \tau$ exhibits a plateau that
would be reminiscent of the Kolmogorov's 4/5 law by 
assuming that $r \propto u_\mathrm{rms} \tau$ and that the Eulerian velocity field remains frozen during the particle displacement.  More precisely, one gets empirically that  
\begin{equation}
\langle \delta u^{(L)}_\parallel (\tau)^3 \rangle \approx - \frac 4 5 \mathrm C~\varepsilon u_\mathrm{rms}\tau
\label{long_3}
\end{equation}
with $C\approx 0.5$ in the range $10< \tau /{\tau_\eta} \lesssim {T_L}/{\tau_\eta}$ at $R_\lambda=280$. 
Let us note that the Kolmogorov's 4/5 law in the Eulerian framework is not observed as clearly for such low $R_\lambda$ \cite{Antonia06}.}

\begin{figure}[t!]
	\centering
	\includegraphics[clip,height=6.3cm,keepaspectratio]{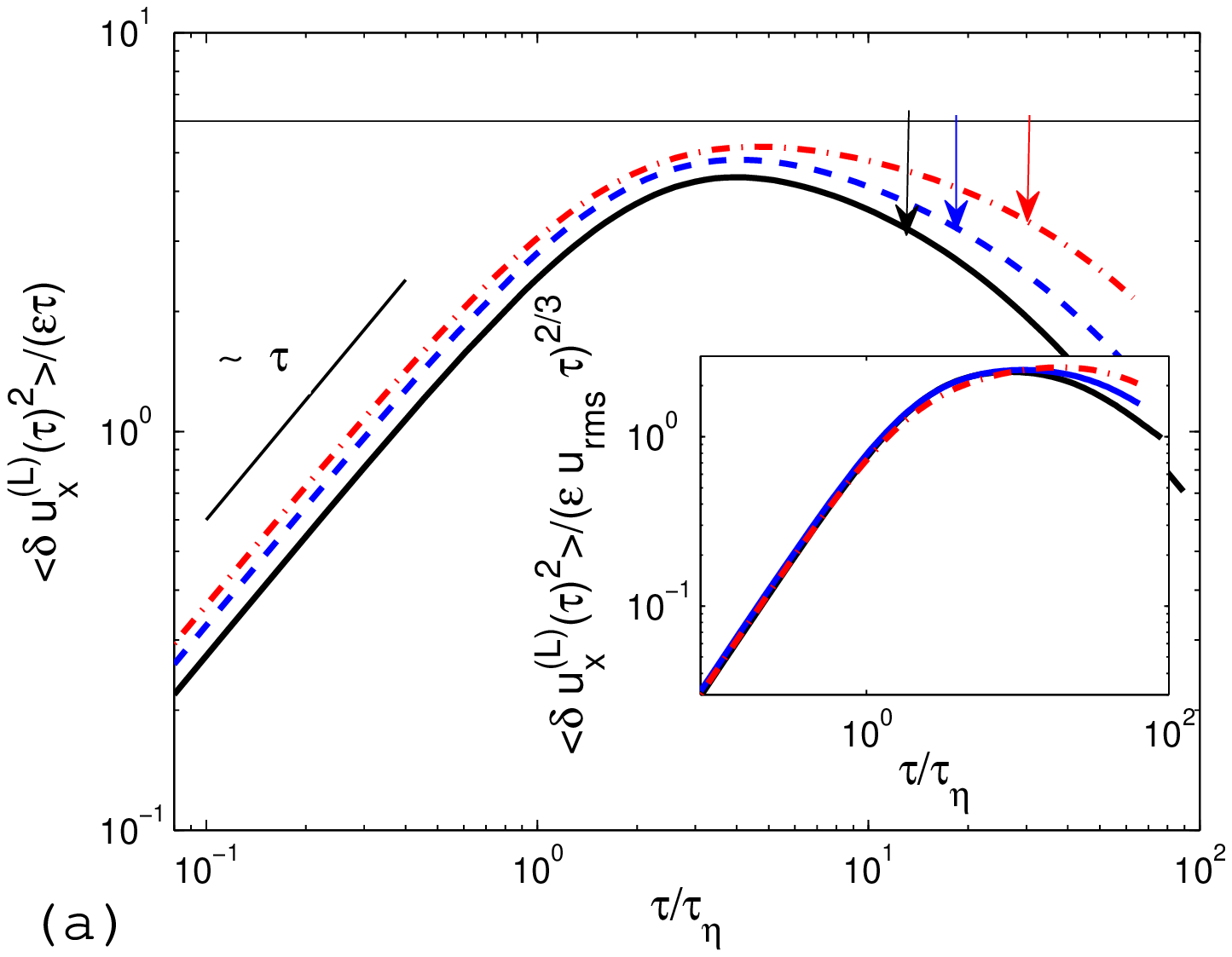}
	\includegraphics[clip,height=6.3cm,keepaspectratio]{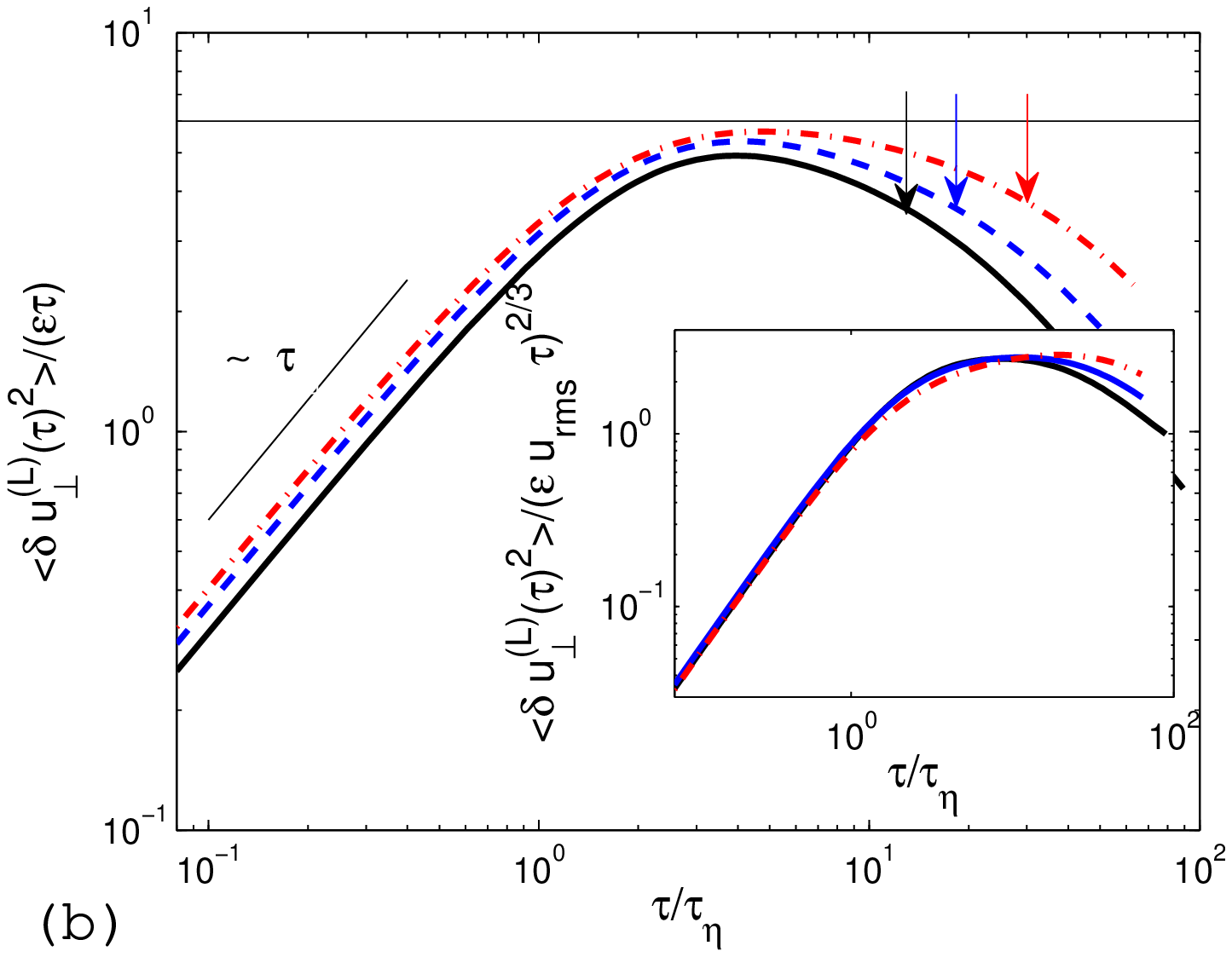}
	\includegraphics[clip,height=6.3cm,keepaspectratio]{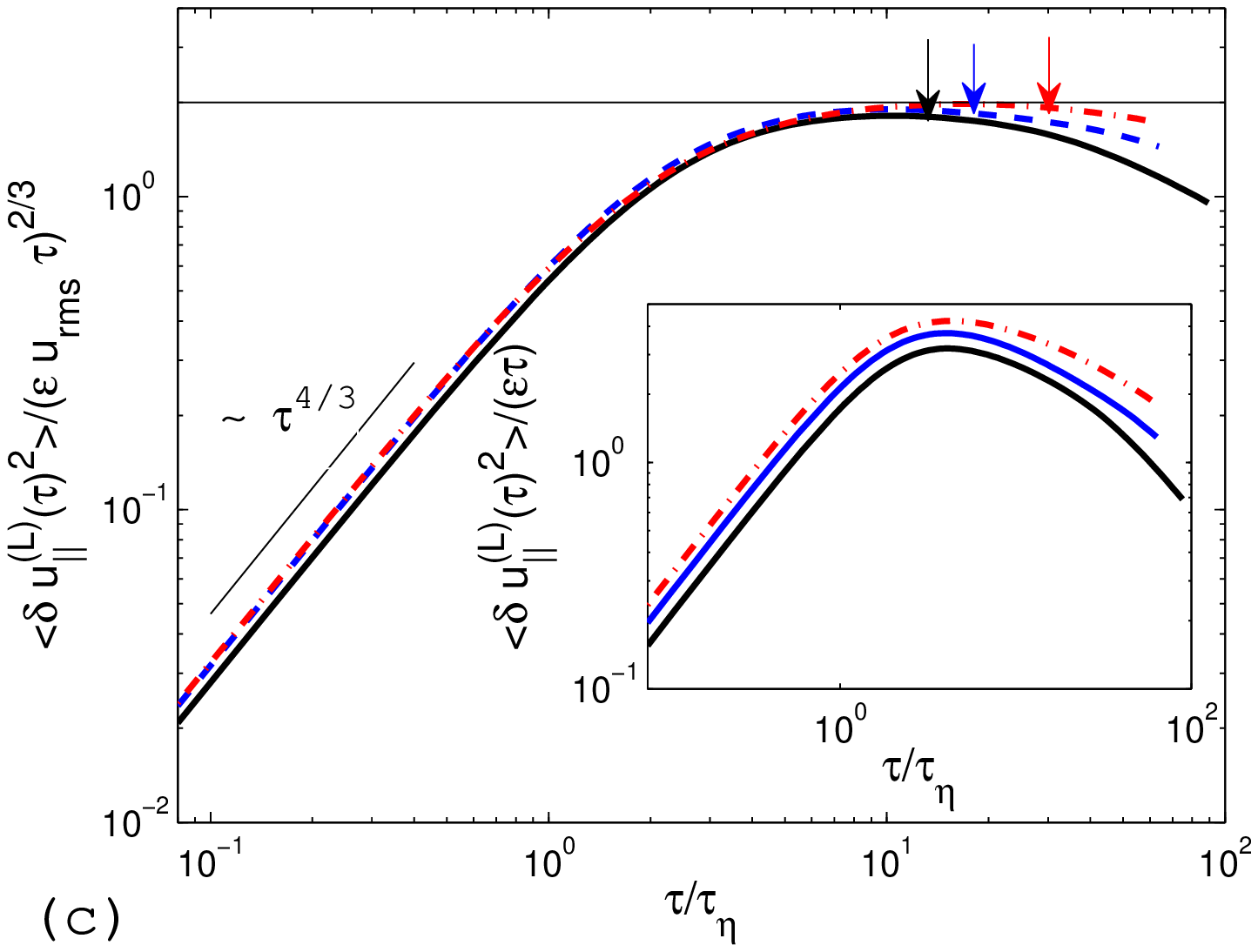}
	\caption{Dependence on $\tau$ of the second-order moment of (a) $\delta u^{(L)}_{x}$, (b) $\delta u^{(L)}_\perp$ and (c) $\delta u^{(L)}_\parallel$, compensated by $\varepsilon \tau$ in (a) and (b), and by $(\varepsilon u_{rms}\tau)^{2/3}$ in (c). In the insets, the moments are compensated by the unintended scaling law.
	The horizontal lines indicate the values $6$ in (a) and (b), and $2$ in (c). The same convention as in Fig. \ref{fig:3} is used for the different $R_\lambda$.}
	\label{fig:2}
\end{figure}

The second-order moment of the different Lagrangian increments are plotted as a function of $\tau$ in Fig. \ref{fig:2}. 
The classical Kolmogorov's phenomenology yields for the second-order moment of the Cartesian increment 
$\langle \delta u^{(L)}_{x} (\tau)^2 \rangle=C_0\varepsilon\tau$ with the constant $C_0\approx 6$ at inertial time scales \cite{LagSF}. 
{Our measurements do not allow us to recover quantitatively this scaling law, as expected at such moderate Reynolds numbers \cite{Lanotte13}, but $\langle \delta u^{(L)}_{x} (\tau)^2 \rangle$ gets closer to the Kolmogorov's prediction as $R_\lambda$ increases. This behavior is consistent with some results already reported in the literature \cite{LagSF}. As already observed, the transverse increment behaves in a similar way as the Cartesian increment (see Fig. \ref{fig:2} (a) and (b)) and, therefore, $\langle \delta u^{(L)}_{\perp} (\tau)^2 \rangle=C_0\varepsilon\tau$ is also expected to hold at inertial time scales (in the limit of infinite Reynolds number).} 
On the other hand, the longitudinal increment exhibits a scaling law compatible with $\langle \delta u^{(L)}_\parallel (\tau)^3 \rangle ^{2/3}$.
By compensating $\langle \delta u^{(L)}_{\parallel} (\tau)^2 \rangle$ by $(\varepsilon u_\mathrm{rms}\tau)^{2/3}$ one infers a plateau at a value  close to $2$ (see Fig. \ref{fig:2}(c)), which eventually leads to
\begin{equation}
\langle \delta u^{(L)}_\parallel (\tau)^2 \rangle \approx 2~(\varepsilon u_{rms}\tau)^{2/3} \label{long_2}
\end{equation}
 in the range $10 <  \tau /{\tau_\eta}\lesssim {T_L}/{\tau_\eta}$ at $R_\lambda=280$.
A direct consequence of Eq. (\ref{long_3}) and Eq. (\ref{long_2}) is  that the skewness coefficient of the longitudinal increment should be constant in the range $10 <  \tau /{\tau_\eta}\lesssim {T_L}/{\tau_\eta}$ with $S^{(L)}_\parallel(\tau)\approx -1/5\sqrt{2} \approx -0.14$  at $R_\lambda=280$, which is indeed verified in Fig. \ref{fig:3}(b).
{Note, however, that  some (weak) dependence of these empirical values on some physical parameters of the turbulent flow is \textit{a priori} not ruled out.}

\begin{figure}[t!]
	\centering
	\includegraphics[clip,height=6.3cm,keepaspectratio]{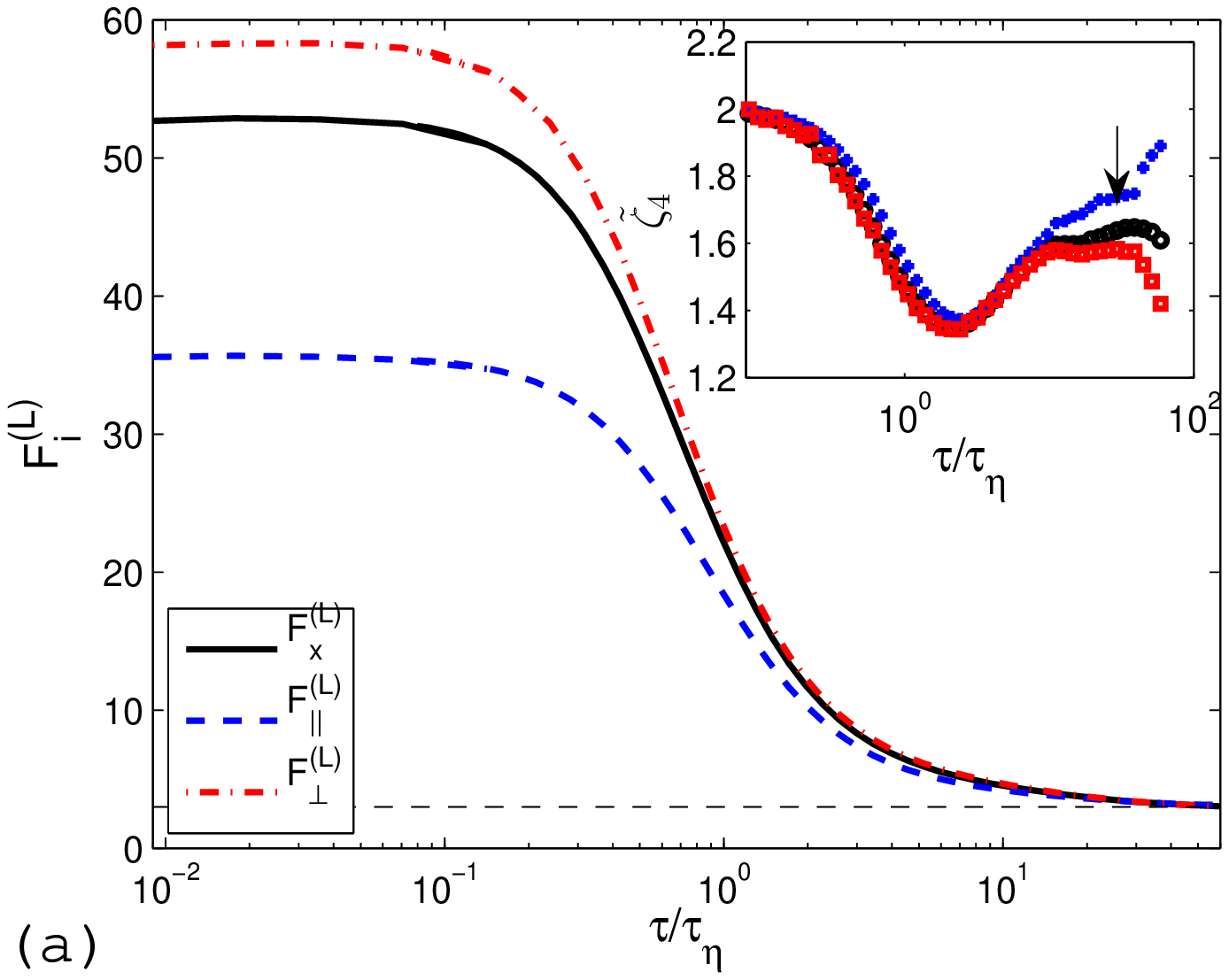}
	\includegraphics[clip,height=6.3cm,keepaspectratio]{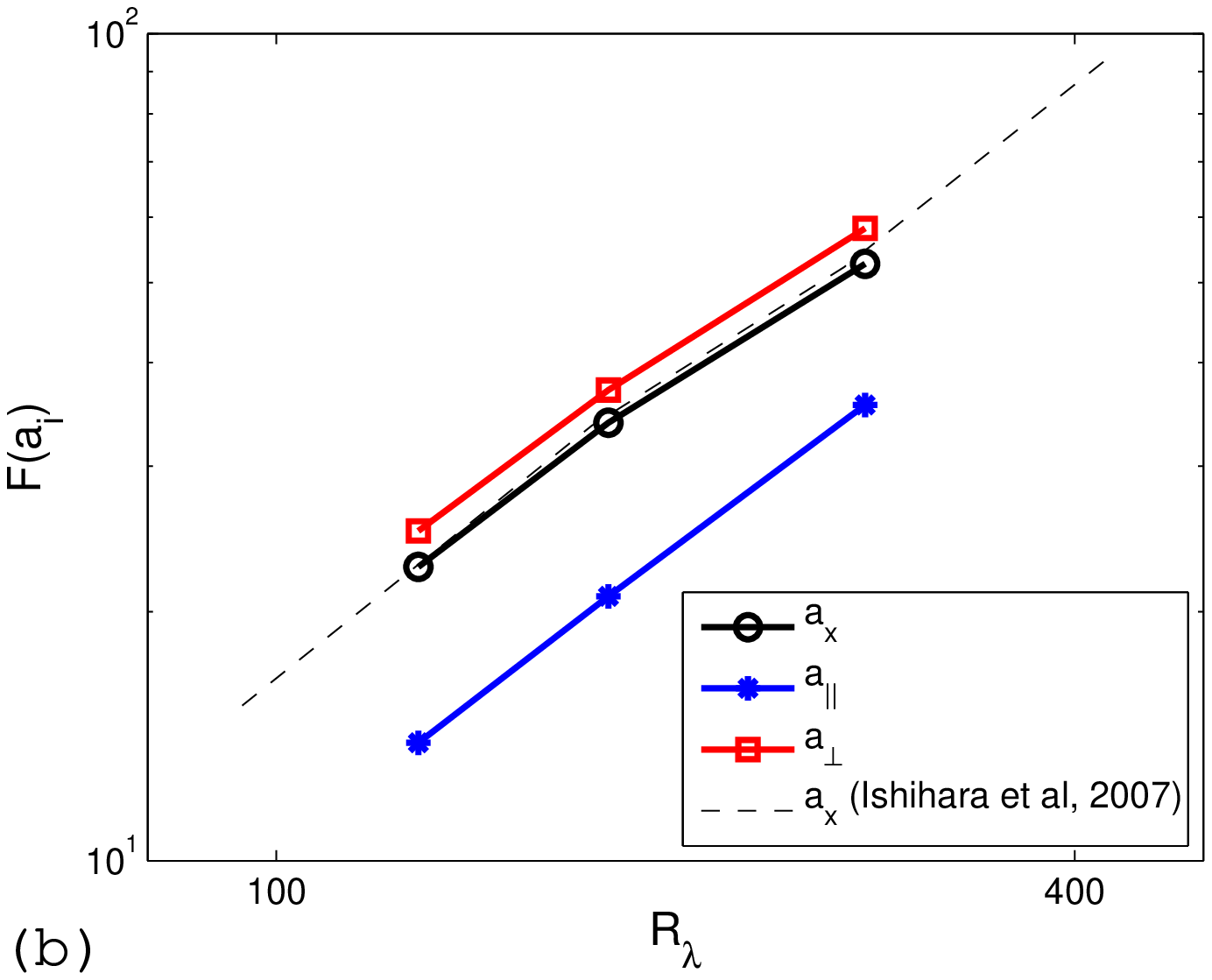}
	\caption{ (a) Flatness of the different Lagrangian increments versus $\tau$ at $R_\lambda=280$ (the dashed line indicates the value $3$ for a Gaussian distribution) --- Inset: Local fourth-order (relative) scaling exponent $\widetilde {\zeta_4}_i=\mathrm d\log\langle\delta u^{(L)}_i(\tau)^4 \rangle / \mathrm d\log\langle\delta u^{(L)}_i(\tau)^2 \rangle$. Squares: transverse increments; Circles: Cartesian increments ; Crosses: longitudinal increments. (b) Flatness of the different acceleration components versus $R_\lambda$. 
	}
	\label{fig:flatn}
\end{figure}

The flatness coefficients $F^{(L)}_i(\tau)= \langle\delta u^{(L)}_i(\tau)^4 \rangle/\langle\delta u^{(L)}_i(\tau)^2\rangle^2$ of the Cartesian, longitudinal and transverse increments are plotted as  functions of $\tau$ in Fig. \ref{fig:flatn}(a). All of them are decreasing functions of the time scale, reflecting a continuous shape deformation of the PDFs from long-tail at vanishing $\tau$ to Gaussian statistics ($F=3$) at large $\tau$ (see Fig. \ref{fig:PDF}(b)). The following ordering is satisfied
\begin{equation}
F^{(L)}_\parallel(\tau)<F^{(L)}_{x}(\tau) \lesssim F^{(L)}_\perp(\tau),\quad \forall \tau>0.
\end{equation}
The main differences between the flatness coefficients occur in the dissipation range ($\tau < \tau_\eta$), where the transverse increment is slightly more intermittent than the Cartesian increment, which is in turn more intermittent than the longitudinal increment.  
Once again, the behavior of $\delta u^{(L)}_{x}$ is closer to the one of $\delta u^{(L)}_\perp$; this has been observed for all the considered $R_\lambda$. 
The  dependence on $R_\lambda$ of the flatness coefficients in the limit $\tau \to 0$ is shown in Fig. \ref{fig:flatn}(b). Surprisingly, all acceleration components exhibit the same dependence in agreement with data from the literature for the Cartesian increment \cite{Ishihara07}.
{In the inset of Fig. \ref{fig:flatn}(a), the local fourth-order (relative) scaling exponent  $\widetilde {\zeta_4}_i=\mathrm d\log \langle\delta u^{(L)}_i(\tau)^4 \rangle  / \mathrm d\log \langle\delta u^{(L)}_i(\tau)^2 \rangle$ is plotted for the three Lagrangian increments. 
The transverse and Cartesian  increments behave quite similarly and agree with the data reported in the review paper \cite{PRL08}. 
Nevertheless, the power-law scaling is more pronounced
for the transverse increment with $\widetilde {\zeta_4}_\perp=1.59\pm 0.02$ in excellent  agreement with experimental data (for the Cartesian increment) at $R_\lambda=1100$ \cite{mordant}.  The longitudinal increment obviously behaves differently at inertial time scales and there is no so clear evidence of power-law scaling; the bottleneck effect \cite{PRL08} seems to propagate deeper in the inertial range.
}

Longitudinal and transverse Lagrangian velocity increments are sensitive to the geometry of fluid trajectories. 
The longitudinal increment  (Eq. (\ref{eq:lagrangian_increment_longitudinal}))  is exactly zero in the case of pure (constant-speed) rotation; its magnitude is expected to be higher when the trajectory is straight, typically when the particle enters a flow-region dominated by a high strain (tending to stretch material fluid elements) and lower when the fluid particle enters a region of high vorticity (tending to spin material fluid elements). On the contrary, the transverse increment (Eq. (\ref{eq:lagrangian_increment_transverse})) vanishes for a straight trajectory and should have a higher magnitude when the trajectory twists itself \cite{braun06,curvature}. This is of course a simplistic view, 
nevertheless, one may consider that longitudinal and transverse Lagrangian increments will be preferentially sensitive to strain and to rotation respectively. To check this feature, we have calculated the variances of the different acceleration components (the flow topology being more naturally defined locally) conditioned on the sign of $\Delta=27R^2+4Q^3$ where $Q=-Tr({\bf m}^2)/2$ and $R=-Tr({\bf m}^3)$ are invariants of the velocity gradient tensor $m_{ab}=\partial_au_b$. This allows us to distinguish strain-dominated ($\Delta<0$) from vorticity-dominated ($\Delta>0$) regions of the flow \cite{Cantwell92}. Our results can be synthesized for the conditional variances as
\begin{equation}
	1<\frac{\langle {a^+_\parallel}^2 \rangle}{\langle {a^-_\parallel}^2\rangle}<\frac{\langle {a_{x}^+}^2 \rangle}{\langle {a_{x}^-}^2 \rangle}<\frac{\langle {a_\perp^+}^2 \rangle}{\langle {a_\perp^-}^2 \rangle},
	\label{ineq:topology}
\end{equation}
with $\langle {a_i^+}^2 \rangle \equiv \langle a_i^2|\Delta>0 \rangle$ and $\langle {a_i^-}^2 \rangle \equiv \langle a_i^2|\Delta<0 \rangle$.
The inequality (\ref{ineq:topology}) indicates that (i) all the acceleration components exhibit stronger fluctuations in vorticity-dominated regions than in strain-dominated ones, and that (ii) strong fluctuations of acceleration along swirling streamlines ($\Delta>0$) are more pronounced for the transverse component than for the longitudinal component of the acceleration. This latter result is in agreement with our expectations. 

\section{Conclusion} Longitudinal and transverse Lagrangian velocity increments have been introduced and examined  in a fluid-mechanical context. 
These increments provide a new path to the characterization of Lagrangian statistics in HI turbulence, and allow us to establish some bridge with  Eulerian statistics. 
Interestingly, the longitudinal and transverse Lagrangian increments exhibit different features. 
The transverse increment is more intermittent and behaves similarly to the standard Cartesian Lagrangian increment. 
By considering their first two statistical moments,
 it is found that  Lagrangian and  Eulerian scalings can be matched by considering the (local) mapping $r \propto u_\mathrm{rms} \tau$ for the longitudinal  increment, whereas the (non-local) mapping $r \propto \langle \delta u_\parallel^{(E)}(r)^2 \rangle^{1/2} \tau$ seems to be more suitable for the transverse  increment. 
%
%
%
Importantly, the skewness coefficient of the longitudinal increment is strictly negative for all time scales, which can be related to time-irreversibility.
%
 In future, state-of-the-art experimental techniques permitting nowadays to get complete three-dimensional traces of fluid-particle trajectories should help to address the robustness of our preliminary results when varying the Reynolds number over a much larger range of values, {and carefully investigate  issues such as intermittency corrections and anomalous scaling laws for these longitudinal and transverse Lagrangian increments}.

This work has benefited from the financial support of the French research agency (grant ANR-12-BS09-0011) and from HPC ressources (PSMN computing center) at the Ecole normale sup\'erieure de Lyon.

\bibliographystyle{unsrt}

\bibliography{longitudinal}





\end{document}